\RequirePackage{textcomp,graphicx}
\documentclass[%
reprint,
showkeys,
amsmath,amssymb,
aps,
prb,
floatfix,
]{revtex4-1}
\usepackage[english]{babel}
\usepackage{graphicx}
\usepackage{amsmath}
\usepackage{amssymb}
\usepackage{natbib}

\usepackage{xcolor,textcomp}
\begin{document}
	
	\title{Real-space observation of the low-temperature Skyrmion lattice in Cu\textsubscript{2}OSeO\textsubscript{3}(100) single crystal}
	
	\author{Gerald~Malsch}
	\author{Peter~Milde}
	\email{peter.milde@tu-dresden.de}
	\author{Dmytro~Ivaneiko}
	\affiliation{Institute of Applied Physics, Technische Universit\"at Dresden, D-01062 Dresden, Germany}
	
	\author{Andreas~Bauer}
	\author{Christian~Pfleiderer}
	\affiliation{Physik-Department, Technische Universit\"{a}t M\"{u}nchen, D-85748 Garching, Germany}
	
	\author{H.~Berger}
	\affiliation{Institut de Physique de la Mati\`ere Complexe, \'Ecole Polytechnique F\'ed\'erale de Lausanne, 1015 Lausanne, Switzerland}

	\author{Lukas~M.~Eng}
	\affiliation{Institute of Applied Physics, Technische Universit\"at Dresden, D-01062 Dresden, Germany}
	\affiliation{Dresden-W\"{u}rzburg Cluster of Excellence -- Complexity and Topology in Quantum Matter (ct.qmat), TU Dresden, 01062 Dresden, Germany}
	
	\date{\today}
	
	\begin{abstract}
		Cu\textsubscript{2}OSeO\textsubscript{3} is a skyrmion host material in which two distinct thermodynamically stable skyrmion phases were identified.
		We report magnetic force microscopy imaging of the low-temperature magnetic phases in bulk Cu\textsubscript{2}OSeO\textsubscript{3}(100) single crystal.
		Tuning the external magnetic field over the various phase transition at a temperature of 10~K, we observe the formation of helical, conical, tilted conical and skyrmion lattice domains in real space.\\
	\end{abstract}
	
	
	
	\maketitle

Skyrmions are topological magnetic excitations which promise great potential for applications in data storage or alternative computational schemes and which have been found in a huge number of materials \cite{Roadmap2020}. Among all the skyrmion host materials Cu\textsubscript{2}OSeO\textsubscript{3} is the only Bloch-skyrmion type magnetoelectric insulator \cite{}. Low Gilbert damping \cite{Stasinopoulos2017} as well as tuning of the skyrmion lattice by electric field \cite{Kruchkov2018, White2018, Huang2018,Wan2020}, strain \cite{Seki2017,Okamura2017, Nakajima2018}, pressure \cite{Deng2020,Crisanti2020,Nishibori2020}, doping \cite{Wilson2019, Sukhanov2019, Neves2020} or geometrical constraints \cite{Wilson2020, Han2020} makes the material an interesting study case in the context of spintronics. Moreover, recently a second low temperature skyrmion phase for magnetic field along the $\langle 100 \rangle$-directions has been reported by small angle neutron scattering, magnetization and specific heat measurements as well as broadband ferromagnetic resonance \cite{Chacon2018,Halder2018,Bannenberg2019,Aqeel2021,Lee_2021}. It was found, that these low-temperature skyrmion lattice state can be stabilized by a special magnetic field protocoll where before a target field value is reached the magnetic field is cycled with small amplitude for several times. Moreover, in contrast to the skyrmion phase close to the Curie temperature which is stabilized by thermal fluctuations \cite{Muehlbauer2009}, this second skyrmion state crucially depends on the cubic anisotropy of the crystal. 
Theoretical discussions highlighted the role of the cubic anisotropy for the existence and orientation of all the various magnetic states that have been observed. With respect to the stability of skyrmions it was predicted, that isolated skyrmions and skyrmion clusters can only exist surrounded by the conical state, while they would disperse in the homogeneous state. This may be different if dense skyrmion lattices form \cite{Leonov2019}.

To date no real-space investigation of the low-temperatue skyrmion state in bulk material has been reported. Here, we show first real-space images of the low-temperature magnetic phases of Cu\textsubscript{2}OSeO\textsubscript{3} using magnetic force microscopy (MFM), which proofed to be a valuable tool when studying spin textures such as helices and skyrmions in Cu\textsubscript{2}OSeO\textsubscript{3} \cite{Milde2016, Zhang2016, Milde2020}. We clearly identify the various magnetic states from the MFM images and investigate the phase transition from the canted conical towards the skyrmion lattice state as well as the transformation of the skyrmion polycrystal into a long range ordered skyrmion single crystal where
the skyrmion lattice orientation becomes locked to the underlying Cu\textsubscript{2}OSeO\textsubscript{3} crystal lattice by magnetic annealing. 

The Cu\textsubscript{2}OSeO\textsubscript{3} single crystal was grown by means of chemical vapor transport, cut from the single-crystal ingot with a wire saw, ground, and carefully polished into a cuboid measuring $1\times1\times1$~mm\textsuperscript{3}.
The orientation and single-crystallinity of the sample was confirmed by X-ray Laue diffraction before it was mounted with conductive silver paste and transferred to the scanning force microscope without further treatment.

\begin{figure*}[t!]
	\includegraphics[width=\textwidth]{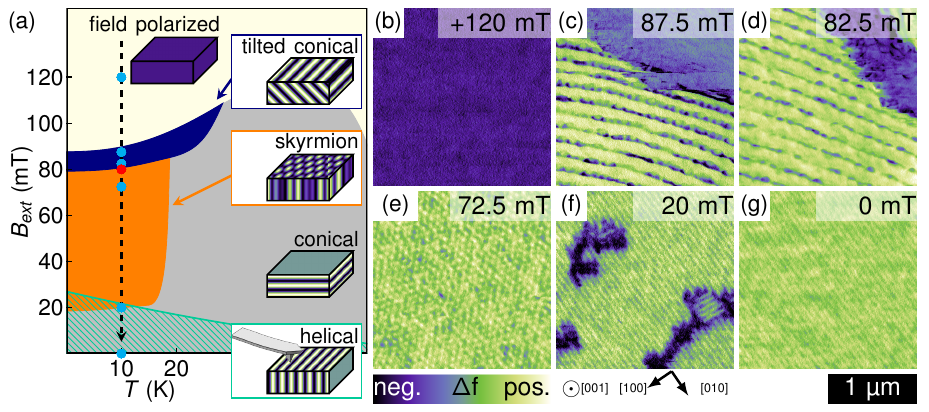}
	\caption{ \label{fig1} Magnetic imaging of a Cu\textsubscript{2}OSeO\textsubscript{3} single crystal in various fields parallel to $\langle 100 \rangle$ at 10K.
		(a) Phase diagram derived from magnetometry for high field cooling and susequent lowering of the field. The helical state with multiple q-vectors, denoted by the turquoise hashed area, is typically only observed after zero field cooling. The insets schematically show the expected MFM patterns.
		Ramping the field down from saturation (b), the conical or field polarized state transforms into the tilted conical state at $B_{ext} = 87.5$~mT (c). In (d) two tilted conical domains are shown. At $B_{ext} = +80$~mT the skyrmion lattice appears (e) which remains to $B_{ext} = +25$~mT. (f) at $B_{ext} = +20$~mT the [100] helical domain intersepted with tiny areas of the [010] helical domain is present. (g) These tiny domains disappear at $B_{ext} = 0$~mT. An image measured in saturation at $B_{ext} = +250$~mT has been used for background correction. The field values of images (b) to (g) are marked by blue dots in (a). The red dot in (a) represents the field $B_{ext} = +80$~mT at which all other measurements presented in figures \ref{fig2} and \ref{fig3} are obtained.
	}		
\end{figure*} 

We performed magnetic force microscopy with an Omicron cryogenic ultra-high vacuum atomic force microscope \cite{Omicron} equipped with the RHK R9s electronics \cite{RHK} using the PPP-QMFMR probes from Nanosensors \cite{Nanosensors} driven at mechanical oscillation amplitudes $A \approx 10$~nm.
Images were recorded at 10~K in a two-step process.
In the first step, the topography and the contact potential difference of the sample was recorded and the topographic 2D slope was canceled.
In the second step, the MFM tip was retracted $\approx 20$~nm to record the magnetic forces while scanning a plane above the sample surface.
After the first MFM image had been completed, the magnetic field was changed automatically in steps of few mT inbetween consecutive images and/or cycled -- i.e. modulated at low frequency and amplitude by sweeping the field in the superconducting magnet -- in order to magnetically anneal the spin textures.
Images for background compensation were taken in the saturated state at $|B_{ext}| = 250$~mT before or after the field sweeps.

In figure \ref{fig1} we summarize a measurement series sweeping the external magnetic field from the field saturated state at $B_{ext} = 250$~mT to zero field as schematically shown in figure \ref{fig1}(a).
The phase diagram is derived from magnetization measurements for high field cooling and subsequent lowering of the field.
The small insets schematically present the expected MFM contrast.
Lowering the magnetic field from the saturated state at $B_{ext} = +250$~mT in figure \ref{fig1}(b), first the conical or field polarized state transforms into the tilted conical state at $B_{ext} = 87.5$~mT. In figure \ref{fig1}(c) a tilted conical domain interfacing to the field polarized or the conical state is shown. A second tilted conical domains separated by a region in the field polarized or the conical state becomes visible in figure \ref{fig1}(d). At a slightly lower field of $B_{ext} = +80$~mT the skyrmion lattice appears, which remains to $B_{ext} = +25$~mT, see exemplarily figure \ref{fig1}(e). At $B_{ext} = +20$~mT the [100] helical domain intersepted with tiny areas of the [010] helical domain appears, shown in figure \ref{fig1}(f). 
Similar population of helical domains with their modulation vector perpendicular to the field axis has been also observed on a (110)-oriented sample before \cite{Milde2020}.
The tiny domains disappear at $B_{ext} = 0$~mT, see figure \ref{fig1}(g). An image measured in saturation at $B_{ext} = +250$~mT has been used for background correction.
All the expected magnetic textures are clearly resolved. Remarkebly, we do not find a stretched skyrmion lattice in the low field region of the phase diagram as it has been recently reported from FMR \cite{Aqeel2021}.

\begin{figure}[tb!]
	\includegraphics[width=\columnwidth]{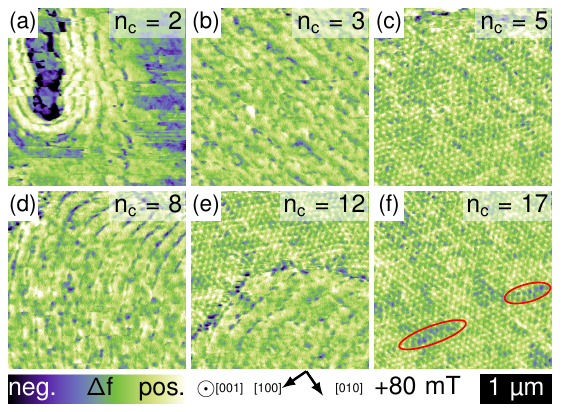}
	\caption{ \label{fig2} Transition from the tilted conical state to the skyrmion lattice state at constant field due to field cycling. 
		(A) In the top left corner a nucleation point of multiple tilted conical domains is visible.
		With just one further field cycling, in (B) a proper tilted conical state has developed in the field of view.
		First skyrmions appear after two more cyclings in (C) being again replaced by another tilted conical domain in (D).
		In (E) both states are present simultaneously. Finally in (F) only the skyrmion lattice is present which then remains also for further field cycling. 
	}	 
\end{figure} 

\begin{figure}[tb!]
	\includegraphics[width=\columnwidth]{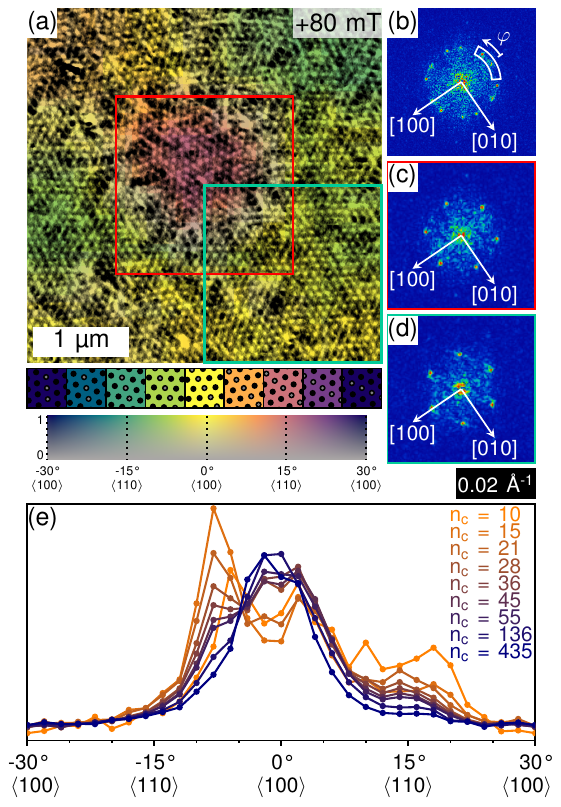}
	\caption{ \label{fig3} Analysis of skyrmion lattice rotation domains.
		(A) MFM image with skyrmion rotation domains. The MFM signal is shown as dark/bright-contrast and the in-plane skyrmion lattice rotation angle is coded by colour. Colour saturation measures the determinateness of the angle assignement. 
		(B) In the FFT of (A) six ring segments with multiple peaks are visible. 
		(C \& D) FFTs of the zoomed areas in the middle and the lower right corner of image (A) show one dominant sixfold peak pattern of the major skyrmion rotation domain which appears to be rotated by $\approx 21$\textdegree ~towards each other.
		(E) As the number of field cyclings increases, the spectral weight of the averaged angular distribution within the ring segment highlighted in (B) condenses into a single peak around $\langle 100 \rangle$ direction due to annealing of the skyrmion crystal locked to the Cu\textsubscript{2}OSeO\textsubscript{3} crystal lattice. 
	}
\end{figure} 

The transition from the multiple-q tilted conical state to the skyrmion lattice state involves a first order phase transition, where mainly two mechanisms for the skyrmion nucleation have been discussed.
In the first mechanism skyrmions are generated by rupture formation of metastable spiral states \cite{Bannenberg2019}.
In the second mechanism torons are involved, which could possibly nucleate at the domain walls between the spiral domains of the tilted conical state \cite{Leonov2020}.
In figure \ref{fig2}, we present an example for such a phase transition by choosing a constant external field of $B_{ext} = 80$~mT but cycling the field inbetween consecutive images with $n_c$ measuring the total number of cyclings.
At this field we found both the tilted conical as well as the skyrmion state in previous experiments.
Coming from the saturated state at $B_{ext} = 250$~mT and directly after setting the new field and two initial field cyclings we find an elliptically shaped pattern in figure \ref{fig2}(a), which we interpret as the nucleus of several tilted conical domains.
Already after just one more cycling in figure \ref{fig2}(b) the whole frame is covered by a proper tilted conical helix.
With further cyclings also the skyrmion lattice appears and dissappears from the field of view in figures \ref{fig2}(c) and \ref{fig2}(d), respectively. Note, that the tilted conical domain in figure \ref{fig2}(d) has a different orientation than that in figure \ref{fig2}(b).  The phase boundary between this tilted conical domain and the skyrmion lattice domain becomes visible in figure \ref{fig2}(e). Overall, we are facing a very dynamic situation, where locally the magnetic texture flips forth and back between these two states.
Finally, in figure \ref{fig2}(f) the skrmion lattice covers the whole frame, and the tilted conical state does not re-appear for further field cycling.
Summing up, we find that both phases coexist in the transition and we conclude that the skyrmion lattice grows on expense of the tilted conical state, which seems to be more related to the rupture formation mechanism then the nucleation mediated by torons. Yet, we note, that the nucleation was not as clearly visible as in the case of Mn\textsubscript{1.4}PtSn \cite{Zuniga2021} or like for inverse process in FeCoSi \cite{Milde2013}. 
Here, only a rather small number of field cyclings was necessary in order to establish the skyrmion lattice, while in general the number will depend on the cycling speed and the field step.

Evenso the skyrmion phase is stabilized here after only $n_c=17$ cyclings, in figure \ref{fig2}(f), still many defect lines within the skyrmion lattice are visible, which hint to the formation of small skyrmion lattice domains forming sort of a skyrmion polycrystal \cite{Rajeswari2015}. Two of the defect lines are marked by red ellipses in the image.
The quite fluctuating MFM-signal makes a direct threshold based identification of the skyrmions and direct analysis of the lattice at least challenging.
Thus, we use an algorithm based on fast Fourier transforms (FFT) of only partial images in order to extract the main orientation of the skyrmion lattice in a certain area. Details on the method can be found in the supplementary information.

We demonstrate this FFT-based skyrmion lattice orientation determination in figure \ref{fig3}(a), where
the MFM signal is shown as dark/bright-contrast, while the in-plane skyrmion lattice rotation angle asigned by our scanned window FFT algorithm is coded by colour. The colour saturation is a measure for the determinateness of the angle assignement. In the center of the image a roughly circular shaped domain nearly locked to a $\langle 110 \rangle$-direction is shown in light red colour. In the bottom right corner a yellow colour indicates a domain locked to  a $\langle 100 \rangle$-direction. This multidomain state is further exemplified by use of Fast-Fourier-transforms shown in figures \ref{fig3}(B to D). In figure \ref{fig3}(B) the transform of the full image is shown, which consists of six ring segments with multiple peaks, whereas in the FFTs of the zoomed areas single sixfold peak patterns are visible as shown in figures \ref{fig3}(C \& D).
In other words, we find rather small skyrmion lattice domains which are characterized by their in-plane rotation angle as a continous order parameter forming a skyrmion polycrystal which should not be confused with the hexatic phase recently reported for a melting skyrmion lattice in Cu\textsubscript{2}OSeO\textsubscript{3} \cite{Huang2020}.
Yet, the cubic anisotropy breaks the rotational symmetry and in consequence locks the skyrmion lattice to high symmetry directions depending on the sign of the cubic anisotropy parameter.
For Cu\textsubscript{2}OSeO\textsubscript{3} this results in a $\langle 100 \rangle$-axes locked skyrmion lattice, whereas for MnSi it would be the $\langle 110 \rangle$-axes \cite{}.
Therefore, we tested if magnetic annealing could transform the skyrmion polycrystal into a proper highly ordered skyrmion crystal, i.e. the skyrmion lattice in its ground state.
In figure \ref{fig3}(E), we show the averaged angular distribution of spectral weight within the ring segment highlighted in figure \ref{fig3}(B). As the number of field cyclings increases, the spectral weight condenses into a single peak around the $\langle 100 \rangle$ direction indicating the annealing of the skyrmion lattice into larger domains locked to the crystal axis by the cubic anisotropy. 

In summary we show first real-space images of the low temperature skyrmion phase in Cu\textsubscript{2}OSeO\textsubscript{3}. Nucleation and ordering of the latter is improved by magnetic annealing. During the annealing an initial coexistence of the skyrmion lattice and the canted conical state can be observed. With longer annealing first a skyrmion polycrystal is formed which then transforms into a long range ordered skyrmion single crystal where the skyrmion lattice orientation becomes locked to the underlying Cu\textsubscript{2}OSeO\textsubscript{3} crystal lattice. We conclude that the thermal activation at these reduced temperatures does not suffice to drive the phase transition at high speed, which explains the large hystereses reported in literature for this transition. At the same time the observed interfaces between different magnetic phases allow to study properties of skyrmions in contact to the conical state. Despite a recent report of FMR-data suggesting a stretched skyrmion lattice in the low field region of the phase diagram \cite{Aqeel2021}, we only find a proper helical phase where all three helical domains become populated.

\section*{acknowledgement}

	P.M., G.M., D.I. and L.M.E.\ gratefully acknowledge financial support by the German Science Foundation (DFG) through the Collaborative Research Center ``Correlated Magnetism: From Frustration to Topology'' (Project No. 247310070, Project C05), the SPP2137 (project no. EN~434/40-1), and project no. EN~434/38-1 and no. MI~2004/3-1. L.M.E. also gratefully acknowledges financial support through the Center of Excellence - Complexity and Topology in Quantum Matter (ct.qmat) - EXC 2147.
	A.B.\ and C.P.\ acknowledge financial support by the Deutsche Forschungsgemeinschaft (DFG, German Research Foundation) under TRR80 (From Electronic Correlations to Functionality, Project No.\ 107745057, Project E1) and the excellence cluster MCQST under Germany's Excellence Strategy EXC-2111 (Project No.\ 390814868) as well as by the European Research Council (ERC) through Advanced Grants No.\ 291079 (TOPFIT) and No.\ 788031 (ExQuiSid) is gratefully acknowledged.


\section*{Supplementary information}

Supplementary information contains details about the computational domain recognition method, additional data measured on a plate sample as well as slideshows of the MFM-image series.

\bibliography{LT-SKX_Cu2OSeO3_Bib}


%
%

%
\newpage

\renewcommand{\thesection}{S\arabic{section}}  
\renewcommand{\thefigure}{S\arabic{figure}} 


	
	
	
	

	
	


\section{Magnetic imaging of plate sample}

\begin{figure*}[tb!]
	\includegraphics[width=\textwidth]{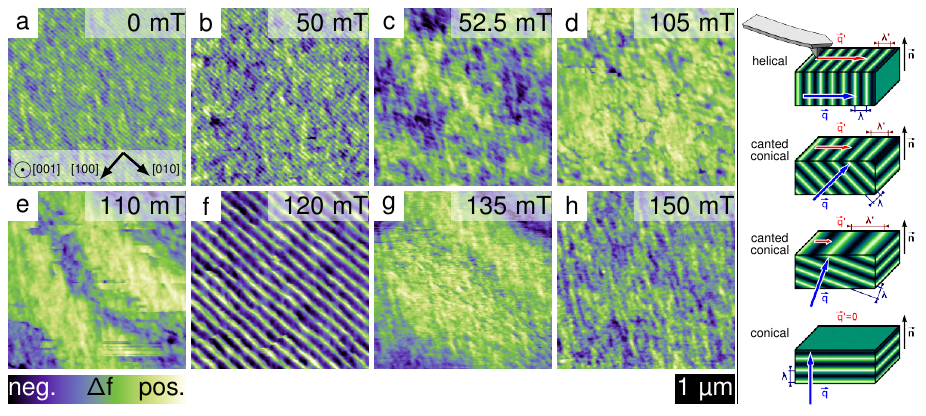}
	\caption{ \label{figS1} Magnetic imaging of a plate sample after zero field cooling in increasing field. Helical and canted conical phase show a distinct 1D-modulated texture. Images are not corrected for topographical background. The color span has been individually adapted for each image. The schematic on the right side illustrates the varying pitch length due to the orientation of the modulated texture  for the different magnetic phases.}
\end{figure*} 

We show a series of images obtained after zero field cooling (zfc) to $T$~=~10~K.
In figure \ref{figS1}a, the pattern of a helical modulation along the [100]-axis is visible, which persists until the external field $B_{ext}$ exceeds a value of $B_{c1,ext}$~=~50~mT (figure \ref{figS1}b).
Above $B_{c1,ext}$~=~50~mT the magnetic images don't show a clear one- or two-dimensional modulated state.
With increasing external field strength also the density of imaging artefacts increases, until they start to resolve into a clearly one-dimensional modulated state with long pitch as signature of the canted conical state at $B_{c2,ext}$~=~110~mT.
For fields above $B_{c2,ext}$ the pitch of the modulation shrinks reaching the minimum at $B_{ext}$~=~122.5~mT, followed by a transition to the saturated state which is reached at $B_{c3,ext}$~=~150~mT.
In summary, the initial magnetization curve starts in the helical phase, passing through the conical state into the canted conical state and ends in the field saturated state.
Running in the same way a full magnetization loop, we find exactly the same magnetic phases and the same critical fields.
Only, in the helical state also the out-of-plane [001] helical domain becomes populated.

Repeating the experiments with magnetic annealing applied after each field step, we see no substantial changes.
Especially, the low-temperature skyrmion phase could not be reached.
Minor changes connected to the magnetic annealing are, first, that the canted conical state develops better in lower field strength, i.e. $B_{c2,ext}$ is slightly shifted towards lower fields, and second, that in some rare cases a domain wall between the out-of-plane [001]-oriented helix and the in-plane [100]-oriented helix was found.
Yet, due to the large size of the helical domains, it is not clear, if the occurance of domain walls in the measurements is provoked by the magnetic annealing, or if it is a statistical effect.
Interestingly, the profiles of the domain walls vary even for one and the same tip and domain wall orientation at the surface.
This already shows, that the underlying 3D-profile and orientation of the wall is not unique, which will be discussed in greater detail in a later publication.

Independent of the application of the magnetic annealing protocol, the observed helical wavelength changes as a function of the external field in both the helical as well as the canted conical state.
A detailed analysis is presented in later.

\section{Analysis of helix director re-orientation}
\label{section:reorient}

\begin{figure}[b!]
	\includegraphics[width=0.5\textwidth]{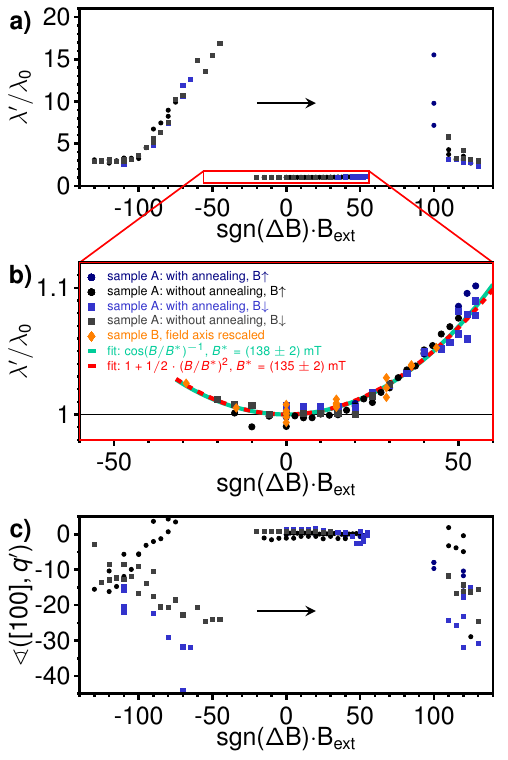}
	\caption{ \label{figS3} Analysis of the apparent wavelength $\lambda^\prime$ and in-plane orientation of the helical director in sample A as a function of external field.
		a) The helical and the canted conical state can be clearly seen. At the low field end of the canted conical state the apparent wavelength increases due to the rotation of the helix into the out-of-plane direction.
		b) Zoom into the field region of the helical phase. Only the in-plane helices are visible to MFM. The apparent wavelength $\lambda^\prime$ varies up to about 10\% for fields close to
		the $B_{c1,ext}$. For comparison also data measured on sample B are displayed. Note, for these data the field axis has been rescaled by the ratio of the demagnetization ratios.
		c) Dependency of the in-plabne angle towards [100]-axis as a function of the external field. While in the helical state always the [100] domain was populated, in the canted conical state the (major) canted conical domain always nucleates at an angle around -15\textdegree.}
\end{figure}

In figure \ref{figS3} we summarize the apparent wavelength $\lambda^\prime$ and the in-plane rotation angle $\varphi = \sphericalangle([100],\boldsymbol{q}^\prime)$ of one-dimensional modulated textures from all measurement series obtained on the plate sample, where $\boldsymbol{q}^\prime$ denotes the component of the helix wave vector in the surface plane.
In order to distinguish hysteresis effects, we inverted the field axis for field sweeps running from positive to negative fields, so that all histories run from the left-hand to the right-hand side.
Also, all data have been scaled to the apparent wavelength of the in-plane helical domain in zero field $\lambda_0 = 59.46 \pm 0.12$~nm.
First we describe the field dependency of the apparent wavelength, presented in figure \ref{figS3}a.
Coming from the field saturated state the canted conical state appears with $\lambda^\prime = 3 \cdot \lambda_0$.
This value is maintained for the next $\approx 30$~mT until the apparent wavelength starts to increase in agreement with a rotation of the helix into the out-of-plane direction.
The maximum size of the scan frame measures $4\times4$~\textmu m, so that the evolution cannot be safely tracked above a value of $\lambda^\prime \approx 18 \cdot \lambda_0 \approx 1$~\textmu m.

When the helical phase is entered, only the in-plane domains ($\boldsymbol{q}\perp[001]$) can be imaged by MFM while the out-of-plane domain ($\boldsymbol{q}||[001]$) delivers no contrast. 
The zoom into the field region of the helical phase in figure \ref{figS3}b reveals a weak field dependency of the apparent wavelength $\lambda^\prime$, which
for fields close to $B_{c1,ext}$ varies by up to 10\%.
For comparison we also add data measured in the helical state on the cuboid sample, yet for these, first the field axis has to be rescaled, according to the different sample shape and demagnetization factors.
In case of the plate sample, the demagnetization tensor reads $\boldsymbol{N} = \boldsymbol{n} \otimes \boldsymbol{n}$ with the surface normal vector $\boldsymbol{n}$.
For the highly symmetric cuboid sample, we use $\boldsymbol{N} = 1/3 \cdot \boldsymbol{E}$ with $\boldsymbol{E}$ being the unity tensor.
Together with the internal suszeptibility tensor of the helix $\boldsymbol{\chi}_{int}(\boldsymbol{Q})$ given by the director of the helix $\boldsymbol{Q} = \boldsymbol{q}/|\boldsymbol{q}|$:
\begin{eqnarray}
\boldsymbol{\chi}_{int}(\boldsymbol{Q}) = \chi^{int}_{||} \cdot \boldsymbol{Q} \otimes \boldsymbol{Q} + \chi^{int}_\bot \cdot ( \boldsymbol{1} - \boldsymbol{Q} \otimes \boldsymbol{Q}) \label{eq3}
\end{eqnarray}
and $\chi^{int}_{||} = 2\cdot\chi^{int}_\bot = 1.76$ deep inside the helimagnetic state \cite{Janoschek_2013},
the resulting scaling factor for in-plane-domains reads $B_{ext,A}/B_{ext,B} = (1+\chi^{int}_\bot)/(1 + 1/3 \cdot \chi^{int}_\bot) = 1.454$.

Apparently, the results obtained on the two quite different samples coincide, which already allows us to exclude both measurement artefacts as for instance a magnetic field dependent scanner calibration as well as effects due to the sample shape.
Thus we conclude, that the observed wavelength field dependency is either intrinsic or a surface effect.

Two scenarios may explain the increasing wavelength in the helical state.
On the one hand, the field dependency of the in-plane helical domains, can be well described by the function $\lambda^\prime(B) = \lambda_0/\cos(B/B^\ast)$ with $B^\ast = (138 \pm 2)$~mT, which implies a linear relation between the out-of-plane angle of the helix director and the external magnetic field strength for a helix with field independent wavelength.
On the other hand, the field dependency can be equally well described by the function $\lambda^\prime(B) = \lambda_0 \cdot \left[1 + 1/2 \cdot (B/B^\ast)^2\right]$ with $B^\ast = (135 \pm 2)$~mT, which is the second order Taylor expansion of the expression above.
The wavelength itself also depends on the external field if the orientation of the helix is fixed perpendicular to the field, and such a dependency would be quadratic in lowest order due to symmetry.
Hence, we cannot decide which of the two scenarios -- rotation or stretching of the helix -- applies. 

Let us now turn back to figure \ref{figS3}a. Leaving the helical state towards higher fields, again no clear one dimensional modulated texture is present, until at around $|B_{ext}| =100$~mT the canted conical state re-appears for another $\Delta B_{ext} \approx 30$~mT.
Thus, we may conclude, that the canted conical state is stable between $100 < |B_{ext}| < 130$~mT, and that we observe a strong hysteresis at the phase boundary between the canted conical and the conical phase, which is related to slow reorientation as well as domain nucleation and annihilation processes.

These reorientation processes are also visible in the evolution of the in-plane angle $\varphi = \sphericalangle([100],\boldsymbol{q}^\prime)$, shown in figure \ref{figS3}c.
While in the helical state always the [100] domain is populated, in the canted conical state the (major) canted conical domain always nucleates at an angle around $\varphi \approx -15$\textdegree, when the canted conical state is reached from high fields.
Lowering the external field then leads to a quite individual rotation of the helical director in the plane for each run.
In clear contrast, the in plane angle of the canted conical domains, when coming from low fields appears to be rather randomly distributed in a region between $-30$\textdegree$\lesssim \varphi \lesssim 0$\textdegree.
Why not the full range $-180$\textdegree$< \varphi < 180$\textdegree ~is covered, remains an open question.

\section{Analysis of skyrmion lattice domains}

In figure 3 in the main paper, we investigated the effect of magnetic annealing on the in-plane skyrmion lattice orientation.
The magnetic field was ramped from high field to $B_{ext} = +80$~mT and between consecutive images at one and the same sample position and field, the external field was shortly swept up and down by $\delta B = 5 \dots 10$~mT.  
From each of the real space MFM images we subtracted a background image obtained in high field.
The resulting difference image was used to compute the two-dimensional Fourier transformation, whereof the absolute values were used. 

First, in order to reduce the effect of the pixelisation of the image on the angular distribution, the Fourier image was transformed into polar coordinates and the values interpolated via bi-cubic splines to achieve a uniform data density in angular direction. 

Second, we removed the noise background. 
The transformed values were summed in angular direction to retrieve a radial intensity distribution. 
By fitting a second-order polynomial to the left and to the right of the k-value corresponding to the skyrmion lattice, a good approximation for the amount of anisotropic noise under the pertinent peaks could be found and subtracted.
Then, the absolute values of the two-dimensional Fourier transformation are summed radially in an interval around the inverse wavelength of the skyrmion lattice. To account for the six-fold symmetry of a triangular skyrmion lattice, the angular distributions are first partitioned into 6 equal intervals and then averaged, resulting in a single-mode distribution over the period of 60 degrees. A sharp peak in the resulting curve corresponds to a uniform skyrmion lattice aligned to a certain direction.

\section{Real space determination of rotational domains}

\begin{figure}[tb!]
	\includegraphics[width=0.5\textwidth]{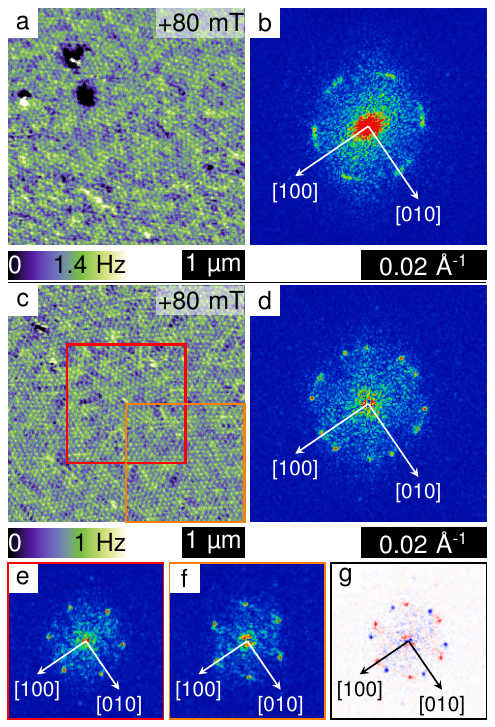}
	\caption{ \label{figS5} Analysis of skyrmion lattice rotation domains.
		a\&c) Typical MFM image showing skyrmions. b\&d) In the fast Fourier transforms six broad ring segments and multiple peaks are visible. 
		e\&f) Zooming into the middle and the lower right corner of image c) and evaluating the corresponding the FFTs. Six peaks are visible in each FFT. 
		g) Calculating the difference between the two FFTs in e)\&f), these sixfold patterns appear to be rotated by $\approx 21$\textdegree ~towards each other.
	}
\end{figure}

\begin{figure}[tb!]
	\includegraphics[width=0.5\textwidth]{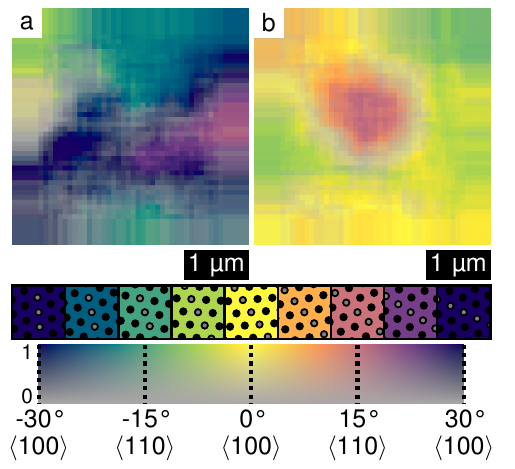}
	\caption{ \label{figS6} Application of our scanned window FFT algorithm to figures \ref{figS5}a\&c.
		The individual rotation domains are labeled by colour, where each colour represents a certain in-plane rotation angle with respect to the [100]-axis.
		Crystallographic equivalent orientations are denoted below the color bar. 
	}
\end{figure} 

For the rotational domain analysis images are processed in a computer-based Fourier analysis using the absolute values of the two-dimensional Fourier transformation. 
The background determined from the Fourier transformation of the entire image in the same way as described above is used to perform the processing on parts of the image to resolve the orientation distribution spatially.
To do so, a frame size smaller than the entire image is chosen, and the Fourier transformation and processing performed only for the data within this frame, centered on a grid of points scanning over the image.
We chose this frame as a square of a fifth of the side length of the entire image, as compromise between the resulting spatial resolution of domains and the angular resolution of their orientation.

As the size of the frame for the Fourier transformation decreases, the amount of pixels in the Fourier image, and therefore the possible resolution decreases. With the presented images and chosen frame size, the ring of peaks corresponding to the skyrmion lattice has a radius of approximately 10 pixels, and thus a circumference of 63 pixels. While the spline interpolation of the data prevents numerical artifacts from the discretization, it cannot increase the amount of information, limiting the angular resolution to the number of pixels along the circumference, which is therefore around 5\textdegree ~for the images presented here.
Again, the angular distributions are first partitioned into 6 equal intervals and then averaged, resulting in a single-mode distribution over the period of 60 degrees.

To display the FFT data spatially, the angular weight distribution of each point is transformed into a complex number and then displayed as a color. 
This number is determined by interpreting the distribution as probability distribution of values along the unit circle in the complex plane and then calculating the mean. 
The argument and absolute of the resulting value within the unit disk, can be interpreted as the predominant direction and intensity of anisotropy, respectively. 

Finally, this complex number is displayed by taking its argument as input for a periodic color scale, giving a color map for its predominant angle, and its absolute as intensity for that color, by linearly interpolating between the latter and a middle gray in the CieCAM color space. 
The appraoch allows the identification of orientation domains at a glance, with areas of uniform lattice appearing as colorful, while areas with multiple orientations or even an amorphous arrangement of skyrmions appear with reduced color saturation or even gray color. 

The result of this moving window FFT algorithm is exemplarily presented in figure \ref{figS6} for the two MFM images shown in figure \ref{figS5}a\&c.
For example, the red area in the center of figure \ref{figS6}b corresponds to one skyrmion lattice domain surrounded by a grayish area, which marks the position of the domain wall to the neighbouring domain.
In the real-space images the sixfold rotation symmetry of the skyrmion lattice restricts the in-plane angles to a range of $\pm 30$\textdegree ~around the $[100]$-axis.
Yet, the relevant potential defining the equilibrium orientation also reflects the crystal symmetry, such that several distinguishable rotational domains are energetically equivalent.


\end{document}